# WANTED
# DEAD OR ALIVE
# EXTRATERRESTRIAL LIFE FORMS

**(Thermodynamic criterion for life is a growing open system that performs self-assembly processes)**

Marko Popovic

Mines ParisTech, CTP, 35 Rue Saint Honoré, 77305 Fontainebleau, France

*"By multiplying the 40x10$^{10}$ galaxies by average stars per galaxy, we have the approximate number of stars in the universe... And so…if only one in a billion of these stars is a "sun" with a planet…and only one in a billion of these is of earth size and composition…there would be something like 3 million worlds with a chance of intelligent life..."*
R. Reagan, U.N. General Assembly, 21 September 1987.

*"Most certainly, some planets are not inhabited, but others are, and among these there must exist life **under all conditions and phases** of development."*
Nikola Tesla

**Abstract:** For more than 100 years, humanity (both specialists and enthusiastic laics) has been searching for extraterrestrial life hoping we are not alone. The first step in the quest for extraterrestrial life is to define what and where exactly to look for. Thus, the basic definition of living matter is a *conditio sine qua non* for the quest. The diversity of species on Earth is so large that our quest for extraterrestrial life cannot be limited to forms and shapes present and known to us from our environment. However, there are two formal conditions that must be fulfilled in order for something to be assumed as living matter. First, it should represent a growing open thermodynamic system (in biological terms - a cell), and thus be a system out of equilibrium. Second, it must perform synthesis, self-assembly and accumulation processes (in biological terms to grow, maintain homeostasis, respond to environment, reproduce, exchange matter and energy, evolve). Populated planets are consisted of two components: biosphere and its environment (geosphere, hydrosphere and atmosphere). Living matter and its environment are out of equilibrium. Thus, the candidate planets are dynamic inhomogeneous systems for two reasons. First, a planet receives energy from its star, which leads to disequilibrium for external reasons. Animate matter contributes to disequilibrium for internal reasons: accumulation of matter and self-assembly. In practice, for a screening, an astrobiologist should search for increase in inhomogeneity on a candidate planet.

1. Introduction:

*"At a workshop in 2003, every member of the International Society for the Study of the Origin of Life was asked to give his or her definition for life, resulting in 78 different answers."* [Palyi 2002]. This may indicate that 78 scientists are searching for 78 different life forms. Obviously, life has not yet been defined clearly, indicating that there are problems in the perception and interpretation of living matter. Defining life requires finding its universal characteristics. All living organisms on Earth share a common ancestor, from which they could have inherited many of the universal characteristics of life on Earth [Domagal-

Goldman 2016]. Even though there is no generally accepted definition of life, there is agreement about the universal properties of life on Earth. The seven properties in common to all known life forms such as ordered cell structure, reproduction, growth and development, energy utilization, response to the environment, homeostasis and evolutionary adaptation [Campbell 2002]. Even though certain non-live phenomena can show some of the seven characteristics of living matter, none can have all of them present at once [Domagal-Goldman 2016]. For example, crystals can grow and have an ordered structure, but do not reproduce, adapt and respond to the environment. Another example are memes, elements of culture that can reproduce and evolve by selection [Dawins 1976; Dennet 1995], but do not have the other characteristics of life.

The most widely accepted definition of life is NASA's working definition that life is "*a self-sustaining chemical system capable of Darwinian evolution*" [Joyce 1994]. This definition distinguishes Darwinian evolution as the most fundamental property of all living organisms and the fact that living organisms are self-sustaining chemical systems. However, it has been criticized. For example, extraterrestrial life may evolve through, not Darwinian, but Lamarckian evolution, in which allele changes appear during an organisms' lifetime [Clealand 2002]. Furthermore, an organism indeed represents a self-sustaining, but even more importantly self-assembled chemical and thermodynamic growing open system out of equilibrium [Boltzmann, Schrodinger, Popovic 2017jbe]. Thus, the NASA working definition appears incomplete.

Lacking a generally accepted definition of life, almost all current attempts to detect life are based on detecting life similar to that on Earth [Plaxco & Gross 2011; Voosen 2017], a criterion that may lead to overlooking many potential forms of life. In fact, both the Viking landers and Curiosity have been looking for (using various methods) fossils or structural components of terrestrial life (amino acids, fatty acids, spores, cells…). The lack of a generally accepted definition of life has led to a more pragmatic approach in all current life-detection experiments: looking for signatures of life similar to those on Earth [Plaxco & Gross 2011]. This includes detection of organic molecules constituting life on Earth, as well as substances consumed and released by metabolism of organisms on Earth [Plaxco & Gross 2011]. The search can be classified into two categories: *in situ* detection using space probes and remote detection using a variety of telescopes [Domagal-Goldman 2016]. *In situ* probes and vehicles are sent to space objects, where they attempt to either directly observe life or find fossils, using cameras and microscopes [Domagal-Goldman 2016]. Furthermore, the probes also look for chemical signatures of life [Davies *et al.* 2009], such as biomolecules like DNA, or products of their decomposition called molecular fossils [Domagal-Goldman 2016].

However, there were attempts to find life through nonequilibrium on exoplanets. These methods are based on analyzing atmospheric spectra, looking for organic molecules that are known not to coexist at equilibrium and are released by life on Earth [Seager 2014; Swain *et al.* 2008]. Other molecular-level evidences are also searched for, such as thermodynamic or kinetic disequilibrium in the environment [Plaxco & Gross 2011]. Since living organisms are systems out of equilibrium [Popovic 2018; Annamalai 2008; Annamalai 2017; Demirel 2010, Balmer 2010], the disequilibrium in organisms leads to disequilibrium in their environments [Domagal-Goldman 2016]. For example, biological activity leads to a gradient of redox species in lake water, which does not exist without life [Domagal-Goldman 2016]. *In situ* detection of life is however limited to our solar system. Remote life detection methods rely on telescopes scanning in various parts of the electromagnetic spectrum. Infra-red and visible spectroscopy can detect presence of molecules known to be produced by life on Earth, like chlorophyll [Domagal-Goldman 2016;

Plaxco & Gross 2011]. Furthermore, living organisms can contribute disequilibrium in the atmosphere, like the simultaneous presence of $O_2$ and $CH_4$, which normally react, so their simultaneous presence implies a constant supply of both compounds by active life forms [Domagal-Goldman 2016]. However, looking for life similar to that on Earth can lead to overlooking many potential forms of life. Therefore, it would be good to have a universally accepted definition of life, capable of encompassing forms of life different than those on Earth.

Sir Arthur Edington [1935] once wrote: "The law that entropy always increases, holds, I think, the supreme position among the laws of Nature. If someone points out to you that your pet theory of the universe is in disagreement with Maxwell's equations — then so much the worse for Maxwell's equations. If it is found to be contradicted by observation — well, these experimentalists do bungle things sometimes. But if your theory is found to be against the second law of thermodynamics I can give you no hope; there is nothing for it but to collapse in deepest humiliation." Moreover, Schrodinger in his book "*What is life?*" chose thermodynamics to define life. Non-equilibrium thermodynamics, developed by Prigogine [Glansdorff 1971], is an excellent tool in analysis of life phenomena [Balmer 2010; Demirel 2014, 2010; Popovic 2017jbe, 2018]. Thus, thermodynamics represents (maybe the most) powerful tool in the extraterrestrial life quest. Thermodynamics should be able to help, because life essentially represents a thermodynamic process [Hayflick 2007plos, 2007nyac, 1998, 2016; Schrodinger 1944, Balmer 2010, Annamalai 2017; Silva 2008; Demirel 2014, 2010; Popovic 2018]. If life is accompanied by aging, and age of living matter represents a biological and thermodynamic state, then life and aging are the change of state [Popovic 2018]. Entropy is the currency of that change [Popovic 2018] So, let's follow the money. Hayflick suggested that entropy (change) explains aging [Hayflick 2007]. If so, then entropy change should also explain life (Popovic 2018arxiv). Then, mass and entropy flow between living matter and its environment. Thus, it should be possible to detect entropy flow between a potential biosphere and its surroundings by detecting disequilibrium and increase in inhomogeneity on a candidate planet.

Every microbiologist from a kilometer distance, using just binoculars, can determine whether a Petri dish contains living organisms. Inhomogeneity in the Petri dish (colonies on agar) will suggest him potential existence of bacteria. By repeated observations, he will see the growth of the colonies and conclude that life not only has existed, but still exists. Without binoculars, he would not be able to confirm existence of bacterial cultures. Thus, lacking technology, it is necessary to rely on theoretical predictions, which will later when the technology is developed be confirmed.

The goal of this paper is to give a definition of life that is wide and can encompass forms of life known on Earth and potential life candidates in the Universe. Section 2 discusses the relationship of biological and thermodynamic criteria for life, to find a wide definition of life. The implications of the thermodynamic definition of life on astrobiology and the search for life and discussed in section 3.

## 2. Thermodynamic foundations of life

The Universe is large enough as Reagan noticed, and more than sufficiently chemically active as Heraclitus noticed claiming "*Panta rhei*", so the probability of abiogenesis and evolution is proportionally large. A wide definition of life in the Universe can be found using thermodynamics, since all properties of life can be translated into very general thermodynamic terms. Boltzmann, a 170 years ago, was the first to introduce entropy (a thermodynamic property) into analysis of living organisms. Schrodinger expanded Boltzmann's consideration. Morrowitz [1992], von Stockar [1999], Hayflick [2016, 2007a, 2007b, 1998], Annamalai [2017], Gladyshev [1999], Glansdorff & Prigogine [1971], Balmer [2010], Garby & Larsen [1995]

and others have emphasized the significance of thermodynamics in explaining life processes. Biological properties in common to all known life forms are: (1) Ordered cell structure, (2) Reproduction, (3) Growth and development, (4) Energy utilization, (5) Response to the environment, (6) Homeostasis, (7) Evolutionary adaptation [Campbell 2002].

Biological properties can be generalized into thermodynamic properties (Table 1): (1a) Represent a thermodynamic system, (2a) Division into subunits, (3a) Accumulate mater and energy- represent growing open system which changes its state, (4a) Exchange energy and matter with its surroundings, (5a) its thermodynamic and thus biological states according to Le Chatelier's principle, (6a) System can be out of equilibrium and maintain homeostasis, for example, an organism represents an open system out of equilibrium because it grows, but each cell maintains homeostasis, (7a) Only the system that successfully adapt to changes in environment keeps integrity and survives.

**Table 1:** Biological characteristics of life and their thermodynamic analogues.

| Biology | Thermodynamics |
| --- | --- |
| Ordered Cell Structure | Represents a thermodynamic system |
| Reproduction | Division into subunits |
| Growth and development | Mass and energy accumulation |
| Energy utilization | Mass and energy exchange |
| Response to the environment | Le Chatelier's principle |
| Homeostasis | Dynamic equilibrium |
| Evolutionary adaptation | Successful adaptation to environment changes, using change in information |

The thermodynamic properties of life (Table 1) should give general guidelines about which kinds of life can be expected in the Universe. Life in the Universe can be best described as open thermodynamic systems (satisfies 1 and 1a) out of equilibrium (satisfies 3, 3a) that perform life processes (satisfies 2-7, 2a-7a). These can be classified into two categories:

a) Growing open thermodynamic systems out of equilibrium without information content, representing the simplest, primitive form of life, on the bottom of the evolutionary scale (primordial vesicles).
b) Growing open thermodynamic system out of equilibrium with information content, which can be at various stages of evolution (prokaryotes, eukaryotes, multicellular organisms).

Both categories can be in both gaseous and liquid phase. Life is impossible only in a solid-state system. The temperature of the system and its surroundings must be above absolute zero. Temperature is an important factor, since it influences kinetics of chemical reactions through the Arrhenius equation.

Every living organism must have cellular structure (represent a thermodynamic system), have a boundary, which separates it from its surroundings and must perform certain chemical processes [Popovic 2018]. The formation of boundaries represents the moment of the formation of a thermodynamic system and abiogenesis [Morowitz 1992]. Also, a living system must be out of equilibrium by exhibiting growth [Popovic 2017jbe]. Thus, it must exchange and accumulate matter and energy with its surroundings.

The abiogenesis and evolution on other planets, except that it must follow the same laws as on Earth, need not have the characteristics, specific form and chemical composition of that on Earth. Formation of life requires a fluid and substances dissolved in it. This is the crucial condition for abiogenesis. On Earth, the fluid is water. On other planets, it can be any appropriate gas or liquid. The most important condition is that the substances dissolved in the fluid have an ability to perform self-assembly: self-organize into a formation that makes a boundary [Morowitz 1992]. On Earth these are phospholipids and triglycerides. For more complex life forms, substances are also necessary which can polymerize into information carrying strings. The polymer would contain information necessary for more efficient development and evolution of such life forms.

Based on the discussion above, the conditions for life are:

1) The most important criterion for life is the general structure – cellular structure separated by permeable boundaries from its surroundings, which represents a growing open thermodynamic system.
2) The system performs synthesis reactions
3) Synthesis products perform self-assembly

Based on these conditions, we suggest a new, more specific definition of life based on NASA working definition: "A self-sustaining, self-assembled and growing open chemical system, capable of Darwinian evolution." Three additional conditions have been added to NASA's definition. It describes living systems more properly. Mathematical formalism for description of living systems should thus be based on nonequilibrium thermodynamics.

## 3. Search for life implications

The thermodynamic conditions for life imply that the criterion for life on other planets is not water, but any fluid capable of dissolving substances that can participate in self-assembly processes. Life need not correspond to our expectances in looks and even phase [Sagan 1976]. Because of these characteristics, such a system can evolve in various directions, which are not necessarily identical or even similar to the directions of evolution on Earth. Therefore, what should be searched for among the planets is not the presence of water, but a fluid as a solvent for molecules which have an ability to react and perform self-assembly processes. In such a fluid, cellular structure (thermodynamic system structure) can appear. Thus, it seems that only cosmic objects that do not contain fluids are not potential candidates for life. It seems that, because of the great width of the term fluid (liquid, gaseous) there are many possible forms of life.

Possible extraterrestrial life might have a form quite different than the one we know on Earth. For example, when surfactants are added to organic solvents, they form structures called inverse micelles. Inverse micelles represent thermodynamic systems, so the first condition is fulfilled. In inverse micelles, the polar heads of the surfactant molecules are oriented towards inside, while the nonpolar tails point towards the organic solvent [Pileni 1993]. Furthermore, if a small amount of water is added, water droplets will form, in which the polar heads will be gathered [Pileni 1993]. Such inverse micelles have been reported to support reactions that are not possible in bulk aqueous solutions [Pileni 1993]. In inverse micelles, reactions were observed like peptide synthesis through enzyme-catalyzed reactions, reaction product extraction, and enhanced-reactivity phenomena [Pileni 1989]. In the described example, the organic-solvent surroundings can contain various organic molecules, like amino acids, nucleosides, fatty acids, glycerol, glucose… Inside such a cell, reaction could occur synthesizing glycolipids, sphingolipids, etc., which could undergo self-assembly into the membrane.

Since compartmentalization binds genotype to phenotype [Morowitz 1992; Plaxco & Gross 2011], membrane self-assembly is a crucial condition for abiogenesis [Morowitz 1992]. Formation of a biothermodynamic system was most likely the key event in abiogenesis [Morowitz 1992]. In the moment when abiogenesis starts, on a potential planet, there is a homogenous thermodynamic proto-system (Figure 1). This corresponds to the prebiotic soup on Earth. By starting synthesis of a substance that has an ability of self-assembly, conditions are made for formation of a border, which divides the newly-formed system from the rest of the proto-system (which now represents the surroundings, or environment, for the newly formed system). Such a scenario has been shown to be possible by Oparin [2003], who experimentally confirmed the possibility of formation of coacervates. A coacervate represents a thermodynamic system, fulfilling the first condition for life. However, it does not fulfill the remaining six conditions. For a coacervate, or an analogous structure on another planet, to become alive, it is necessary for it to start the synthesis process inside itself. The products of the synthesis reaction would have to spontaneously undergo self-assembly into more complex forms (a membrane), which would accumulate in the cell. In that way, the system would have the six characteristics of primitive life.

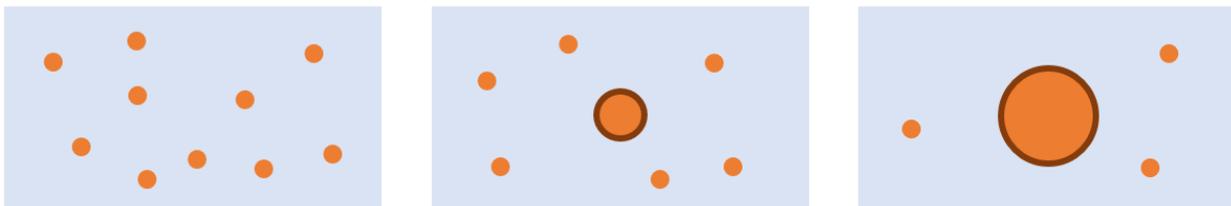

**Figure 1:** Abiogenesis: Homogenous system (orange balls are dissolved substances, blue is water) -> Heterogenous system (the dark brown circle is the membrane, separating the living system from its environment) -> growing biosphere within the planet. Astrobiologist should search for inhomogeneous space objects. The inhomogeneity should increase in time.

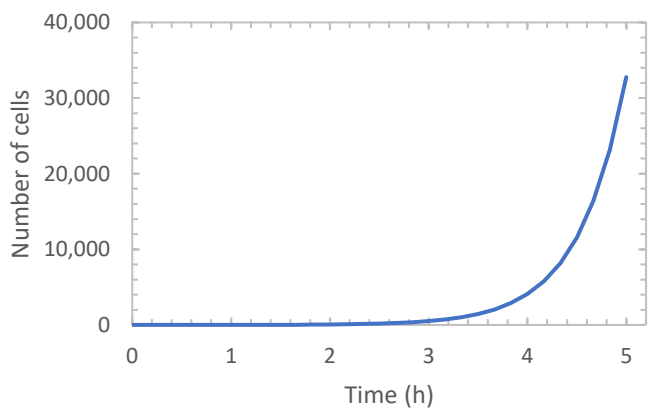

**Figure 2:** Increase in the number of cells as a function of time, using the microbial growth equation [Hogg 2005]. The mass of the cells increases proportionally to their number.

So, what we should be looking for are inhomogeneities on planets containing fluids. The example above leads to the conclusion that extraterrestrial life must have the same general structure (cell), but need not have the same form as life on Earth.

Living cells double their mass every 20 min. to 20 hours (Figure 2). On the other hand, planets are characterized by approximately constant mass. A planet consists of a biosphere and its environment (geosphere, hydrosphere and atmosphere). Thus, the increase in biosphere mass leads to decrease in environment mass. So, in practice we should search for the increase of inhomogeneity on a candidate planet. In a similar way, our microbiologist was looking for inhomogeneity in a Petri dish (bacterial colony).

If bacteria were seeded on an adequate medium on an adequate planet (terraforming), which is uninhabited, then we could monitor the change of the newly formed biosphere mass, as well as the mass of its environment. The result of this experiment is expected to be identical to planting bacteria in a petri dish. The petri dish, like a planet is a closed system, so the bacterial colony growth is followed by a proportional decrease in mass of the agar. This disbalance can be detected visually (fig.1).

Since life can have various forms and use a variety information storage media, evolution could be manifested in various ways not similar to those on Earth. Thus, it might be effective first to look for thermodynamic signs of life and then for evidence of evolution. Information can be contained in nucleic acids and proteins. However, information can also be stored in imperfect crystals [Popovic 2017] and any polymer consisted of aligned asymmetric atoms and molecules (i.e. H, CO…) in an information string [Popovic 2015]. The forms of information storage, and life in general, may vary, but the principles of organization are always the same. Information can be contained in strings of various kinds of material carriers of information. It is not necessary for a string to contain amino acids or nucleotides. Information theory allows the information carrier to be any asymmetric atom or molecule. So, depending on conditions on the candidate planet, the information string can consist of various information carriers.

Thus, our search for extraterrestrial life should be based on the principles of thermodynamics and information theory because both corresponds to the biological principles. This greatly widens the span of candidates for extraterrestrial life. Therefore, the primary search should be for the thermodynamic and informatic signs of potential living matter, followed by the search for Darwinian evolution.

**4. Conclusions**

An expansion of NASA's definition of life has been suggested, including self-assembly processes, growth, as well as the fact that living organisms are open thermodynamic systems. The enormous number of space objects and extensive chemical activity in the Universe results in great probability of existence of extraterrestrial life on various stages of chemical and Darwinian evolution across the Universe. The driving force, physical/chemical laws and structure of life are the same on each planet, but the phenotype forms could be completely different from those on Earth. Looking for extraterrestrial life, we should look for time changing inhomogeneities on potential candidate planets. An ideal candidate should contain a fluid(s), be inhomogeneous, divided into a thermodynamic system (a vesicle or cell like structure) and its surroundings, the system should perform chemical activity, self-assembly and accumulation of matter (growth). These criteria should extend the quest for life to many more candidates.